\documentstyle[aps,epsfig,multicol]{revtex}
\draft
\input epsf
\begin{document}
\title{Microscopic mechanisms of thermal and driven diffusion of non rigid 
molecules on surfaces}
\author{C. Fusco\thanks{Author to whom correspondence should be
addressed. Electronic address: fusco@sci.kun.nl.} and A. Fasolino}
\address{Department of Theoretical Physics, University of Nijmegen,
Toernooiveld 1, 6525 ED Nijmegen, The Netherlands}

\maketitle

\begin{abstract}

The motion of molecules on solid surfaces is of interest for technological 
applications such as catalysis and lubrication, but it is also
a theoretical challenge at a more fundamental level.
The concept of activation barriers is very convenient for the 
interpretation of experiments and as input for Monte Carlo simulations but
may become inadequate when mismatch with the substrate and molecular 
vibrations are considered.   
We study the simplest objects diffusing on a substrate at finite 
temperature $T$, namely an adatom
and a diatomic molecule (dimer), using the Langevin approach.
In the driven case, we analyse the characteristic curves, comparing the 
motion for different values of the intramolecular spacing, both for $T=0$ 
and $T\ne 0$. The mobility of the dimer is higher than that of
the monomer when the drift velocity is less than the natural stretching 
frequency. The role of intramolecular excitations is crucial in this 
respect.
In the undriven case, the diffusive dynamics is considered as a function of
temperature. Contrary to atomic diffusion, for the dimer it is not 
possible to define a single, temperature independent,
activation barrier. Our results suggest that vibrations can account for 
drastic variations of the activation barrier. This reveals a complex
behaviour determined by the interplay between vibrations and a temperature 
dependent intramolecular equilibrium length. 

\end{abstract}

\begin{multicols}{2}

\section{Introduction}
\label{sec:intro}

The diffusion of adatoms and molecules on surfaces is recently attracting 
much attention, in order to understand many properties of technological 
interest, such as thin film growth, catalysis and dissociation~\cite{b.Gomer,b.Kellogg1}. Useful experimental techniques for probing the surface 
diffusion of molecular adsorbates have been developed. 
These allow to follow the dynamical details
of surface diffusion, using Scanning Tunneling Microscopy~\cite{b.Ganz}, 
and to image physisorbed atoms and clusters, using 
Field Ion Microscopy~\cite{b.Kellogg2}.
Most theoretical works have focussed on the determination of energy barriers
for diffusion in different systems~\cite{b.Raut,b.Sholl,b.Linderoth,b.Schunack,b.Boisvert}, usually on the basis of energy arguments and neglecting the 
role of internal degrees of freedom. However, it is often suggested that 
diffusion dynamics can be strongly affected by the presence of 
intramolecular motion~\cite{b.Krylov,b.Hamilton}.
Here we address this problem for the simple but important case of a dimer
physisorbed on a periodic substrate, trying to link the macroscopic 
diffusive behaviour to microscopic degrees of freedom. 
Our theoretical model contains
the salient features of a diatomic molecule physisorbed on a periodic
substrate. However, our work does not aim to probe diffusion in a specific 
molecule-surface system, but to understand diffusion mechanisms expected for
systems of this kind.

In Sec.~\ref{sec:model} we give some details of the model we have used. 
Sec.~\ref{sec:driven} is devoted to the discussion of the results in the 
presence of an external driving, while Sec.~\ref{sec:diffusion} 
deals with pure thermal diffusion. 
Some concluding remarks are presented in the last section.

\section{Model}
\label{sec:model}

We consider a monomer (adatom) and a dimer (diatomic molecule) moving on a 
periodic one-dimensional substrate at finite temperature. 
The particle-substrate interaction is modelled by a periodic function $U$.  
Specifically, for the monomer
\begin{equation}
U(x)=U_0(1-\cos(kx))
\end{equation}
and for the dimer
\begin{equation}
U(x_1,x_2)=U_0(2-\cos(kx_1)-\cos(kx_2)),
\end{equation}
where $x$ represents the spatial coordinate and $k=2\pi/a$, $a$ being the 
lattice constant of the substrate.

For the dimer the interparticle interaction is taken to be 
harmonic: 
\begin{equation}
V(x_1,x_2)=\frac{K}{2}(x_2-x_1-l)^2,
\end{equation}
where $l$ is the spring equilibrium length.
In order to take into account the finite temperature $T$ of the substrate, 
we adopt the Langevin approach.
In this framework the motion of the sliding object is described exactly, while
the substrate is treated as a thermal bath. 
The equations of motion of the monomer and the dimer are respectively
\begin{equation}
\label{e.mon}
m\ddot{x}+m\eta\dot{x}=-U_0\sin(kx)+f+F
\end{equation}
and 
\begin{equation}
\label{e.dim}
\left\{ \begin{array}{ccc}
m\ddot{x}_1 +m\eta\dot{x}_1 & = & K(x_2-x_1-l)-U_0\sin(kx_1)+f_1+F \\
m\ddot{x}_2 +m\eta\dot{x}_2 & = & K(x_1-x_2+l)-U_0\sin(kx_2)+f_2+F,
\end{array} \right.
\end{equation}
where an explicit damping term $m\eta\dot{x}_{i}$ modelling energy 
dissipation has been introduced ($\eta$ is the phenomenological friction 
coefficient), and the effect of finite temperature $T$ is taken into account 
by the stochastically fluctuating forces $f_i$. These two terms are related by
the dissipation-fluctuation theorem:
\begin{equation}
\label{e.flucdiss}
<f_i(t)f_j(0)>=2mk_BT\delta_{ij}\delta(t).
\end{equation}
It is convenient to rewrite the equation of motion in
adimensional units by introducing a characteristic time
$$
\tau=\left(\frac{m}{k_BTk^2}\right)^{1/2}
$$
and defining
$$
\tilde{x}=kx,\quad \tilde{t}=t/\tau,\quad \tilde{\eta}=\eta\tau,\quad 
\tilde{U}_{0}=U_0/(k_BT)
$$
$$
\tilde{f}=f/(kk_BT),\quad 
\tilde{F}=\nolinebreak F/(kk_BT),\quad \tilde{l}=kl,
\quad \tilde{K}=K/(k^2k_BT).
$$
For typical values $m\sim 2\cdot 10^{-26}Kg$, $T=300$K, $a\sim 2\AA$, we 
have $\tau\simeq 0.25ps$. 
In this way Eqs.~(\ref{e.mon}) and (\ref{e.dim}) become 
(in the following we omit the tildes for simplicity)
\begin{equation}
\label{e.monadim}
\ddot{x}+\eta\dot{x}=-U_0\sin x+f+F
\end{equation}
and 
\begin{equation}
\label{e.dimadim}
\left\{ \begin{array}{ccc}
\ddot{x}_1 +\eta\dot{x}_1 & = & K(x_2-x_1-l)-U_0\sin x_1+f_1+F \\
\ddot{x}_2 +\eta\dot{x}_2 & = & K(x_1-x_2+l)-U_0\sin x_2+f_2+F,
\end{array} \right.
\end{equation}
and the fluctuation-dissipation relation Eq.~(\ref{e.flucdiss})
\begin{equation}
<f_i(t)f_j(0)>=2\eta\delta_{ij}\delta(t).
\end{equation}
We perform Molecular Dynamics (MD) simulations, integrating the equations 
of motion using a velocity-Verlet algorithm, with 
time step $\Delta=10^{-4}\tau$, averaging the trajectories over several 
hundreds of realizations ($\simeq 300$ in driven case and $\simeq 3000$ in 
the undriven case with $F=0$), in order to reduce the statistical noise due 
to the stochastic term.

\section{Driven case}
\label{sec:driven}

\begin{figure}
\epsfig{file=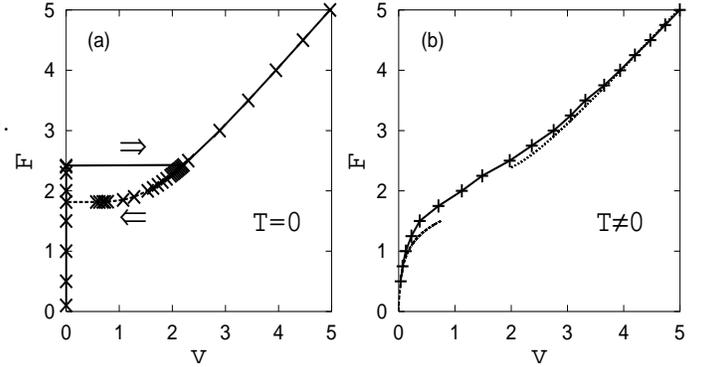, width=9cm, height=7cm}
\caption{Characteristic curve of the monomer for $T=0$ (a) and $T\ne 0$ (b).
The arrows in (a) indicate the parts of the curve where the force is 
increased or decreased. The lines with crosses in (a) and (b) are the 
numerical data, while the dashed line and the dotted line in (b) are fits 
to the data respectively according to Eq.~(\ref{e.vlow}) and 
Eq.~(\ref{e.vhigh}). The parameters used are $U_0=2.5$ and $\eta=1$.}
\label{vFmon.eps}
\end{figure}

\begin{figure}
\epsfig{file=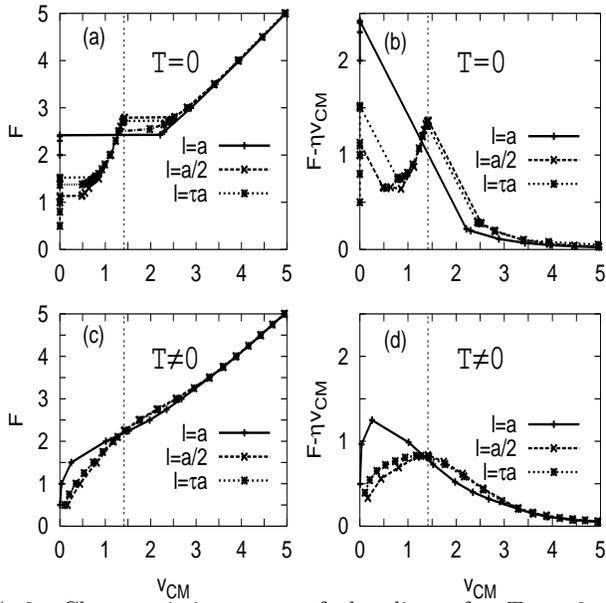, width=9cm, height=8cm}
\caption{Characteristic curves of the dimer for $T=0$ (a),(b) and $T\ne 0$ 
(c),(d) for three values of $l$. The vertical dot-dashed line passes
through $v_{CM}=\omega_0$. The parameters used are $U_0=2.5$, $\eta=1$ and 
$K=1$.}
\label{vFdim.eps}
\end{figure}

\begin{figure}
\epsfig{file=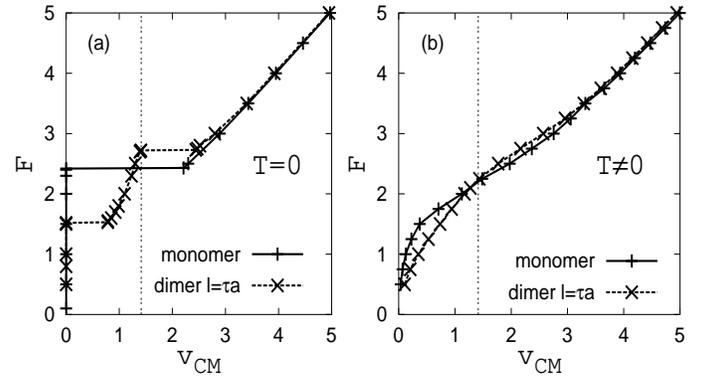, width=9cm, height=7cm}
\caption{Comparison between the characteristic curves of the monomer and 
the dimer for $T=0$ (a) and $T\ne 0$ (b). 
The parameters used are the same as in Figs.~\ref{vFmon.eps} and 
\ref{vFdim.eps}.}
\label{vFmon-dim.eps}
\end{figure}

We have analyzed the response properties of Eqs.~(\ref{e.monadim}) and 
(\ref{e.dimadim}) by studying the characteristic curves, i.e. the behaviour 
of the mobility of the system subjected to the external force $F$.
For the monomer, we can rely on the results of Kramers' Transition State
Theory (TST)~\cite{b.Kramers,b.Hanggi,b.Risken}, yielding analytical 
results for certain regimes of friction and for relatively high values of 
the diffusion barrier $2U_0$. We present in Fig.~\ref{vFmon.eps} the 
results obtained with our simulation for $U_0=2.5$ and $\eta=1$, 
which we will use as reference in discussing the behaviour of the dimer. 
Fig.~\ref{vFmon.eps}(a) shows the velocity-force characteristic in the 
$T=0$ case. Increasing $F$ adiabatically the velocity remains
zero until a critical force $F^{mon}_1$ is reached, since the system has to 
overcome a finite static friction force in order to get out of the 
potential well. 
Then the particle slides with constant velocity $v$ on the substrate, 
resulting in a linear relation between $v$ and $F$ for high values of the 
driving force:
\begin{equation}
F\simeq\eta v.
\end{equation}
If we decrease the force adiabatically a hysteresis effect is 
observed, since the mobility is different from zero also when $F<F^{mon}_1$ 
and vanishes at $F=F^{mon}_2<F^{mon}_1$. 
This corresponds to a bistability between the locked
and the running solution, which is present when the friction coefficient 
$\eta$ is not very large, $\eta/\sqrt{U_0}<1.19$~\cite{b.Risken}.
In the case $T\ne 0$ this bistability disappears, at least for 
intermediate-large values of $\eta$. The system has always the 
chance to get untrapped because of thermal fluctuations. As a consequence 
the mobility is always different from zero when $F\ne 0$,
as shown in Fig.~\ref{vFmon.eps}(b). If $\eta$ is relatively
large, it is possible to describe analytically the behaviour for small 
$F$ by means of Kramers' TST~\cite{b.Kramers}. The result
for the behaviour of $v(F)$ is
\begin{eqnarray}
\label{e.vlow}
v & = & \left\{\left[\frac{\eta^2}{4}+U_0\cos
\left(\arcsin\left(\frac{F}{U_0}\right)\right)\right]^{1/2}-
\frac{\eta}{2}\right\}\times
\nonumber\\
& & {}\exp(-2U_0)\left(\exp(Fa/2)-\exp(-Fa/2)\right)
\end{eqnarray}
As we can see from Fig.~\ref{vFmon.eps}(b), Eq.~(\ref{e.vlow}) reproduces 
the simulation results for $F<1$, but this range increases for larger 
values of $U_0$, where the activated processes are more pronounced.  
The behaviour of the mobility for large $F$, where  
the particle performs a drift motion with a small contribution of the noise 
term can be shown to be~\cite{b.Persson}
\begin{equation}
\label{e.vhigh}
F=\eta v\left(1+\frac{U_0^2}{2v^4}\right).
\end{equation}
In particular, if $F$ is very large compared to $U_0$ the 
term in parenthesis tends to one and the characteristic curve becomes 
linear:
\begin{equation}
F\simeq \eta v.
\end{equation}
Now we consider the driven dimer described by Eq.~(\ref{e.dimadim}). 
In this case, the presence of the interparticle interaction renders 
the problem more complex, giving rise to a richer dynamical behaviour. 
We have analyzed the characteristic curves for different values of the 
intramolecular spacing $l$,
namely $l=a$, $l=a/2$ and $l=\tau_g a$ where $\tau_g=(1+\sqrt{5})/2$ is the 
golden mean, both for $T=0$ and $T\ne 0$ (see Fig.~\ref{vFdim.eps}), and 
compared them to the monomer characteristics (see Fig.~\ref{vFmon-dim.eps}).
Note that we have chosen $l=\tau_g a$ because it is a prototypical 
example of an incommensurate length ratio.

For the dimer it is convenient to rewrite the equation of motion using the
centre of mass (CM) and relative coordinates, defined by
\begin{equation}
x_{CM}=(x_1+x_2)/2\qquad y_r=x_2-x_1-l.
\end{equation}
From Eq.~(\ref{e.dimadim}) we obtain 
\begin{equation}
\label{e.xcmyr}
\left\{\begin{array}{lll}
\ddot{x}_{CM} & = & -\eta\dot{x}_{CM}-U_0\cos((y_r+l)/2)\sin x_{CM}+F\\
\ddot{y}_{r} & = & -\eta\dot{y}_r-2Ky_{r}-2U_0\sin((y_r+l)/2)\cos x_{CM}
\end{array}\right.
\end{equation}
In our MD simulations we choose the initial configuration which minimizes
the total potential energy.
As for the monomer a critical force $F^{dim}_1$, which depends on the value 
of $l$, is needed to achieve motion for $T=0$.  
Then for larger values of $F$ the velocity increases as a function of the
external force, but at a certain value of the force $F^{dim}_3$  
another plateau appears in the $v_{CM}-F$ plane ($v_{CM}$ being the CM 
mean velocity), signalling a dynamical crossover in the system. 
Finally, keeping on increasing the force, the linear regime 
is recovered (Fig.~\ref{vFdim.eps}(a)). 
In the CM frame, the external potential leads, for a drift motion 
$x_{CM}\sim v_{CM}t$, to a time-periodic force acting on the particles, 
with ``washboard'' frequency given by $v_{CM}$. 

The force $F^{dim}_3$, where the second plateau appears, physically 
corresponds to the point where the washboard frequency is in resonance with
the stretching frequency of the dimer $\omega_0=(2K)^{1/2}$, exciting the 
internal degrees of freedom. This resonance 
mechanism was also found in the Frenkel-Kontorova model in the low friction
limit~\cite{b.Strunz}. 
For the dimer two regions of bistability can be observed, 
corresponding to the two plateaus in the characteristic curve; thus, when
$F$ is decreased a first hysteresis occurs in proximity of 
$v_{CM}=\omega_0$, giving a critical value $F^{dim}_4<F^{dim}_3$, 
where the characteristic curve has a discontinuous derivative, 
while a second hysteresis is found when 
$F$ is decreased further from $F^{dim}_1$ and another critical value 
$F^{dim}_2<F^{dim}_1$, where $v_{CM}=0$, is obtained as for the monomer
(these hysteresis curves are shown only for the $l=\tau_g a$ dimer in 
Fig.~\ref{vFdim.eps}). For a closer comparison between the monomer and the 
dimer at $T=0$ see Fig.~\ref{vFmon-dim.eps}(a). 
We note that the qualitative behaviour is the same for different $l$
but the values of $F^{dim}_{1}$ and $F^{dim}_2$ can differ significantly
as a function of $l$. 
This is due to energetic reasons: the dimer with $l=a/2$ is energetically
favourite since, on average, it has to overcome a lower barrier, while for
$l=a$ the two particles tend to be pinned in the minima and to behave like
a monomer (in fact the characteristic curve of the $l=a$ dimer is 
practically superimposed on that of the monomer). 
The resonance frequency is highlighted 
in Fig.~\ref{vFdim.eps}(b) by plotting $F-\eta v_{CM}$
as a function of $v_{CM}$: a clear peak of the characteristic curves 
at $v_{CM}=\omega_0$ is visible when $l\ne a$.

\begin{figure}
\epsfig{file=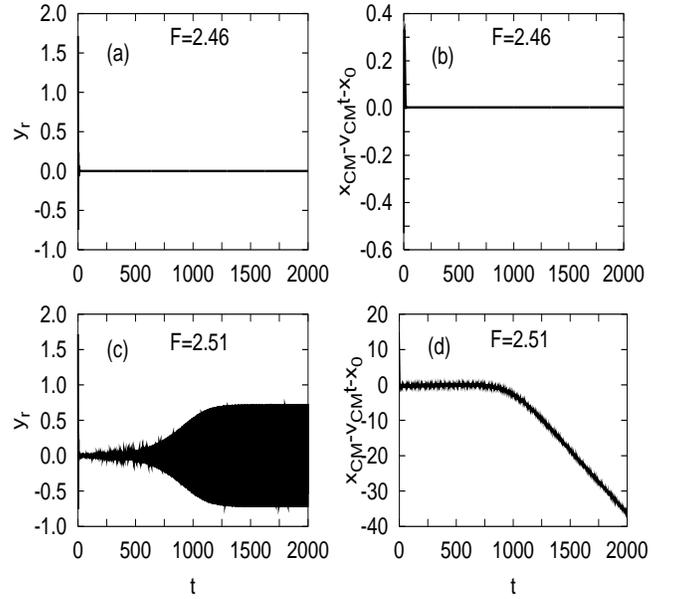, width=9cm, height=8cm}
\caption{Relative and CM motion of the dimer for $l=a$ with initial 
condition $x_2-x_1=1.27a$, for different values of the external driving 
$F$. $y_r$ is shown in (a) and (c), while the deviation of 
$x_{CM}-x_0$ from $v_{CM}t$ is plotted in (b) and (d) ($x_0$ 
is the initial condition for CM).
The parameters used are the same as in Fig.~\ref{vFdim.eps}.}
\label{yrdetl1.eps}
\end{figure}

In order to understand better the resonance mechanisms we consider
the case $l=a$ at $T=0$. 
Assuming a CM drift motion, 
$x_{CM}(t)=x_0+v_{CM}t$, and linearizing the equation of 
motion~(\ref{e.xcmyr}) for $l=a$ we obtain the following equation for $y_r$:
\begin{equation}
\label{e.parametric}
\ddot{y}_r+\eta\dot{y}_r+2Ky_r=U_0\cos(v_{CM}t+x_0)y_r.
\end{equation}
This is the equation of a parametric oscillator, for which an exponential 
increase of the amplitude is expected for $v_{CM}\simeq 2\omega_0$ 
within a given instability window, which we estimate to be
$2.21<v_{CM}<2.37$. We note that indeed the amplitude of $y_r$
increases exponentially in a certain range of $F$, as shown in 
Fig.~\ref{yrdetl1.eps}, but it saturates at long times. 
This is due to the fact that Eq.~(\ref{e.parametric}) assumes a constant 
CM velocity. Actually, in the full system Eq.~(\ref{e.xcmyr}) $x_{CM}$ is 
coupled to $y_r$, so that $v_{CM}$ decreases slightly during the dynamics as
shown in Fig.~\ref{yrdetl1.eps}(d). 
This is enough to shift $v_{CM}$ out of the instability
window, thus stopping the increase of $y_r$.
  
This shows how intramolecular vibrations can be resonantly excited due to 
the sliding on a periodic substrate and that the details of the 
resulting relative motion are non trivial. 
For instance, whether this could represent a 
mechanism for dissociation depends on the maximum excursion from the
equilibrium distance. Moreover, the very nature of parametric resonances 
makes the temporal behaviour very much dependent on the initial values of 
the interatomic spacing which is in turn related to vibrational energy and
temperature.

The behaviour outlined for $T=0$ smears out at finite temperatures. 
Although a hysteretic behaviour has been found for long periodic chains in 
the low friction limit~\cite{b.Braun}, no bistabilities and hysteresis are 
present in our case, as shown in Fig.~\ref{vFdim.eps}(c). 
The static friction force vanishes and the mobility 
is sensitive to the value of $l$. In particular the dimer with $l=a$ has 
the lowest mobility. Note that the curves for different $l$ still cross at 
$v_{CM}=\omega_0$. This crossover of the mobility can be observed also 
comparing the monomer and the dimer (Fig.~\ref{vFmon-dim.eps}(b)).
Moreover, as it can be seen from Fig.~\ref{vFdim.eps}(d), plotting 
$F-\eta v_{CM}$ vs. $v_{CM}$, the resonance peak at $v_{CM}=\omega_0$ still 
survives for $l\ne a$. For $l=a$ the peak is shifted to a smaller value of 
$v_{CM}$. This is due to the fact that in the commensurate case,
even though the static friction force found at $T=0$ vanishes, a larger 
force is needed to reach the sliding state and this corresponds to the 
point of highest curvature in the characteristic plotted in 
Fig.~\ref{vFdim.eps}(c).
To a certain extent this resembles the monomer case, where a very similar 
behaviour is found.

\section{Diffusion}
\label{sec:diffusion}

\begin{figure}
\epsfig{file=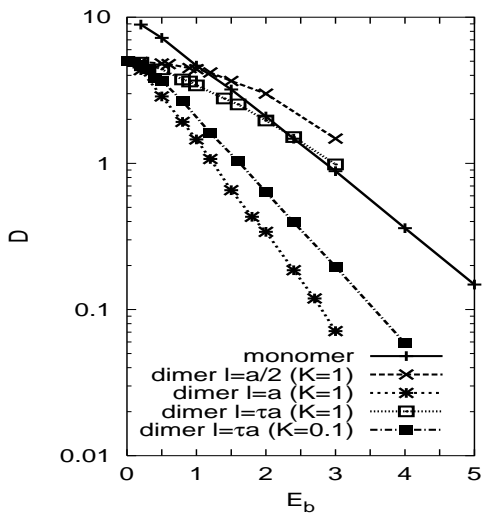, width=7cm, height=7cm}
\caption{Diffusion coefficient of the monomer and the dimer for different
values of $l$ and $K$ vs. $E_b=2U_0$. Here $F=0$ and $\eta=0.1$.
Note that for the free diffusion case $E_b\rightarrow 0$ the diffusion 
coefficient of the monomer is $\frac{1}{\eta}=10$ while that of the dimer 
is $\frac{1}{2\eta}=5$ (independent of $l$).}
\label{diffus.eps}
\end{figure}

Next we examine the pure thermal diffusion, i.e. we study the motion of the 
monomer and the dimer with $F=0$. This problem is 
computationally more time consuming, since the motion is completely random 
and an averaging over many MD realizations is needed. We extract 
information about the diffusive behaviour of the particles by computing the 
Mean Square Displacement (MSD) $<x^2(t)>$, where $<\cdot>$ denotes the 
average over the realizations. The diffusion coefficient $D$ is defined by
\begin{equation}
D=\lim_{t\rightarrow\infty}\frac{<x^2(t)>}{2t}. 
\end{equation}  
Usually it is assumed that the dependence of $D$ on temperature should 
follow the Arrhenius law:
\begin{equation}
\label{e.diffusion}
D=D_0\exp(-E_a)
\end{equation}  
where $D_0$ is a prefactor and $E_a$ is the activation energy for 
diffusion, scaled to $k_BT$. In Eq.~(\ref{e.diffusion}) both $D_0$
and $E_a$ are assumed to be $T$ independent. 
However, some recent studies have already shown that there may be 
deviations from the Arrhenius behaviour~\cite{b.Krylov,b.Hamilton,b.Montalenti}.
Fig.~\ref{diffus.eps} shows the diffusion coefficient $D$ as a function
of the single-particle energy barrier $E_b=2U_0$ (scaled to $k_BT$)
for the monomer and the dimer for different values of $l$ and two values of 
$K$. We note that while the monomer has a uniquely defined activation 
energy $E_a\simeq 0.95$, a clear deviation from the Arrhenius law is 
observed in the dimer case, except for the case $l=a$, 
where it is less evident (as explained in Sec.~\ref{sec:driven} 
the commensurate dimer dynamics is similar to that of the monomer).
The $l=a/2$ dimer diffuses faster than the monomer (at least for $E_b>1$)
due to the most favourable energy configuration (see also~\cite{b.Sholl}). 
Note that the crossover point in the diffusion behaviour can depend on the 
force constant $K$, as it can be seen by comparing the two curves for 
$l=\tau_g a$ with $K=1$ and $K=0.1$.
The dependence of the activation energy on temperature 
(or equivalently on the energy barrier) can be attributed to a temperature 
dependent intramolecular length, which affects the diffusion behaviour 
especially in the high temperature regime (low $E_b$). A similar mechanism
was also proposed in a model of heteroepitaxial island 
diffusion~\cite{b.Hamilton}. Since we are considering small values of $E_b$ 
(rigorously Eq.~(\ref{e.diffusion}) should be valid for 
$E_b\rightarrow\infty$), finite-barrier effects are also important, 
as illustrated in Ref.~\cite{b.Montalenti}.
We are currently trying to unravel the latter from the effects due to 
vibrations. Our results suggest that the
role of intramolecular vibrations is relevant for the diffusion dynamics.
The possibility to excite internal degrees of freedom, even for a simple
diatomic molecule, can affect the temperature dependence of the activation
energy~\cite{b.Krylov} and determine a complex diffusion behaviour which 
can depend on the lattice commensurability and on the interatomic spacing.
We plan to address this issue more extensively in a future work.

\section{Conclusions and perspectives}
\label{sec:conclusion}

We have presented a simplified model to study the driven and undriven motion
of monomers and dimers on a periodic substrate. We have pointed out the  
peculiar effects of the substrate on the CM and relative motion of the dimer,
such as nonlinear mobility, bistabilities and resonance processes.
In this respect, the coupling between the internal degrees of freedom is
crucial. In particular, the diffusion dynamics
reveals strikingly complex features determined both by energetic mechanisms 
and by the role of intramolecular motion. It would be worthwhile to explore
further this issue going from the simple $1D$ model to a more realistic 
approach, considering for example the motion and the orientation of large 
molecules on a $2D$ surface and the diffusion of long chains.

\begin{acknowledgements}

This work was supported by the Stichting Fundamenteel Onderzoek der Materie
(FOM) with financial support from the Nederlandse Organisatie voor 
Wetenschappelijk Onderzoek (NWO).

\end{acknowledgements}

\end{multicols}

\end{document}